\documentclass[11pt,a4paper]{article}
\usepackage[utf8]{inputenc}
\usepackage{amsmath}
\usepackage{amsfonts}
\usepackage{amssymb}
\usepackage{tikz} 

\newcommand{\dt}[1][t]{\,\mathrm{d}{#1}}	
\newcommand{\bracket}[1]{\left( {#1} \right)}	
\newcommand{\der}[2][]{\frac{\mathrm{d} {#1} }{\mathrm{d}{#2}}}
\newcommand{\sder}[2][]{\frac{\mathrm{d}^2 {#1} }{\mathrm{d}{#2}^2}}
\newcommand{\partialder}[2][]{\frac{\partial {#1} }{\partial{#2}}}
\newcommand{\spartialder}[3][]{\frac{\partial^2 {#1} }{\partial{#2}\partial{#3}}}
 		
\newcommand{\lagr}{\mathcal{L}}
\newcommand{\ham}{\mathcal{H}}

\let\oldref\ref
\renewcommand{\ref}[1]{(\oldref{#1})}
\renewcommand{\epsilon}{\varepsilon}
\let\oldint\int
\renewcommand{\int}{\oldint\limits}

\begin{document}
\title{Fermat's Principle in General Relativity via Herglotz Variational Formalism}
\author{Joanna Piwnik\thanks{joanna.piwnik@edu.uni.lodz.pl}, Joanna Gonera \thanks{joanna.gonera@uni.lodz.pl}, Piotr Kosiński \thanks{piotr.kosinski@uni.lodz.pl}}
\date{}
\maketitle
\begin{center}
Faculty of Physics and Applied Informatics\\
University of Lodz, Poland
\end{center} 

\begin{abstract}
New form of Fermat's principle for light propagation in arbitrary (i.e. in general neither static nor stationary) gravitational field is proposed. It is based on Herglotz extension of canonical formalism and simple relation between the dynamics described by the Lagrangians homogeneous in velocities and the reduced dynamics on lower-dimensional configuration manifold. This approach is more flexible as it allows to extend immediately the Fermat principle to the case of massive particles and to eliminate any space-time coordinate, not only $x^0$.
\end{abstract}

\section{Introduction}
Light propagation in gravitational field is described by the solutions to the Maxwell equations in curved space-time. In the short wavelength limit the geometric optics approximation can be used. The basic notion is here that of light rays. In the case of constant gravitational field, their trajectories are described
by Fermat's principal. 

The application of Fermat's principle appears to be particularly useful in the studying of gravitational lensing \cite{30, 31, 32, 33}. One can then assume that the gravitational field is constant since it varies significantly on time scales much larger than the light crossing time. 

Fermat's principle for static space-times has been derived in many papers and books (see, for example, \cite{15,16,17}) and then extended to the stationary case \cite{18,19,20}. The derivation presented in \cite{20} is particularly nice since it combines the stationary phase approximation and equivalence principle. 

However, the lensing phenomena in non-stationary gravitational fields (cosmic strings \cite{34}, gravitational waves \cite{35}) may also become important. Therefore, it seems profitable to generalize the Fermat principle to the case of arbitrary space-time metric. Such a generalization has been proposed by Kovner \cite{21} and studied in detail in \cite{22,23,24}. An important contribution to the subject was provided by Frolov \cite{25} who applied Pontryagin's minimum principle to obtain an effective Hamiltonian for null geodesics in a curved space-time. 

In the present paper we propose an alternative formulation of Fermat's principle in arbitrary gravitational field. To this end we address to the Herglotz variational principle \cite{1,2,3,4,5,6,7,8,9,10,11,12,13}. 

From the formal point of view the light propagation in gravitational field can be described in terms of geometry of null geodesics. Once the affine parameterization of the latter is given they provide the solutions to the Lagrange equations resulting from a simple Lagrangian $H$ quadratic in derivatives. Our derivation of Fermat's principle is based on the following simple statement. Given an autonomous nondegenerate Lagrangian which is homogeneous in generalized velocities one can eliminate one generalized velocity using the energy condition; then the resulting dynamics of remaining coordinates is described by the Herglotz variational principle with the generalized Lagrangian identified with this velocity. This prescription, when applied to the case of constant gravitational field, implies that one computes $\dt[x]^0$ from
the condition $\dt[s]^2=0$ and uses $\int\dt[x]^0$ as the action functional; then the Fermat principle reads $\delta \int\dt[x]^0 = 0$. For the arbitrary gravitational field it agrees with the results discussed in \cite{21,22,23,24}, yielding an alternative formulation of Fermat's principle. 

The general result concerning homogeneous Lagrangians can be applied to the propagation of massive particles as well. As a result one obtains the counterpart of Fermat's principle for such
particles (cf. also \cite{22}). 

The paper is organized as follows. In section 2 we describe in some detail the Herglotz formalism since it is not very familiar. We sketch alternative derivations of some results, including slightly more general form of Noether's theorem. Section 3 is devoted to the discussion of dynamics described by the Lagrangians homogeneous in generalized velocities. It is shown that the reduced dynamics, obtained by eliminating one variable, is described by Herglotz principle. This result is then used in Section 4 to derive
the general form of Fermat's principle valid for arbitrary gravitational field. We show that one obtains a new formulation of Fermat's principle, quite convenient in applications. In Section 5, the application to the Schwarzshild-deSitter metric in comoving coordinates is discussed. We derive the equations describing the light rays trajectories in these coordinates and find the local integral of motion following from the Noether theorem. Using the latter we show that the relevant dynamics is immediately integrable.

Finally, short Section 6 is devoted to some conclusions.
\section{Herglotz variational principle}

The basic notion of analytical mechanics is that of Lagrange function, $L=L(q,\dot{q},t)$ (all equations written out below
correspond to one degree of freedom; the generalization to more degrees is straightforward). Once $L$ is given, the dynamics is described by the variational principle. For any curve $q=q(t)$ in configurational space one defines the action functional
\begin{equation}
\label{1}
S[q] \equiv \int^{t_1}_{t_0} L \bracket{q(t), \dot{q}(t), t} \dt + S_0
\end{equation}
with $S_0$ being an arbitrary fixed constant. The dynamical evolution in the interval $t_0 \leq t \leq t_1$ is obtained by demanding the stationarity of $S[q]$ with respect to the variations $q(t) \to q(t) + \delta q(t)$ with fixed ends
\begin{equation}
\label{2}
\delta S[q] = 0, \hspace{5mm} \delta q (t_0) = 0 = \delta q (t_1)
\end{equation}
As a result, one obtains Euler-Lagrange equations of motion.

Alternatively, the above variational principle may be formulated as follows. For any given curve $q = q(t), \hspace{2mm} t_0 \leq t \leq t_1$, one solves the differential equation
\begin{equation}
\label{3}
\dot{S} = L (q(t), \dot{q}(t), t)
\end{equation}
together with the initial condition
\begin{equation}
\label{4}
S(t_0) = S_0
\end{equation}
The physical trajectory $q(t)$ is obtained from the variational principle
\begin{equation}
\label{5}
\delta S(t_1)=0, \hspace{5mm} \delta q (t_0) = 0 = \delta q (t_1)
\end{equation}
Once put in such a form the variational principle may be readily generalized \cite{1,2,3,4}. Namely, one considers the
generalized Lagrange function, depending also on the action variable,
\begin{equation}
\label{6}
L = L(q,\dot{q}, t,S)
\end{equation} 
The initial value problem \ref{3}-\ref{4} is replaced, for
any curve $q=q(t),\ t_0 \leq t_1$, by
\begin{align}
\label{7}
\dot{S}(t) &= L \bracket{q(t), \dot{q},t,S}\\
\label{8}
S(t_0) &= S_0
\end{align} 
The physical trajectory is obtained from the variational principle
\begin{equation}
\label{9}
\delta S(t_1)=0, \hspace{5mm} \delta q(t_0) = 0 = \delta q (t_1)
\end{equation}
It is not difficult to show that the Euler-Lagrange equation(s) is replaced by the following ones \cite{4}
\begin{align}
\label{10}
&\der[S]{t} - L \bracket{q, \dot{q}, t,S} = 0\\
\label{11}
&\partialder[L]{q} - \der[]{t} \bracket{\partialder[L]{\dot{q}}} + \partialder[L]{S} \partialder[L]{\dot{q}} = 0
\end{align}
The Herglotz variational principle is particularly useful in description of nonconservative systems \cite{5,6}.
 
Herglotz formalism shares many advantages of standard Lagrange formalism. Euler-Lagrange equation determine the relevant Lagrangian up to a total derivative of an arbitrary function of generalized coordinates and time. Let us look for the counterpart of this property in the case of Herglotz principle. The physical information is contained in the trajectory $q=q(t)$. Therefore, let us consider the following transformation
\begin{align}
\label{12}
q'=q\\
\label{13}
t'=t\\
\label{14}
S'=S'(q,t,S), \hspace{5mm} \partialder[S']{S} \neq 0
\end{align}
Inverting eq.\ref{14} one obtains
\begin{equation}
\label{15}
S = N(q,t,S')
\end{equation}
Let us define the new Lagrangian \cite{7,8}
\begin{equation}
\label{16}
L'(q,\dot{q}, S') \equiv \bracket{\partialder[S']{t} + \partialder[S']{q} \dot{q} + \partialder[S']{S}L}_{S \to N(q,t,S')}
\end{equation}
It is straightforward to check that the equations \ref{10}, \ref{11} are equivalent to
\begin{align}
&\der[S']{t} - L'(q,\dot{q}, t,S')=0\\
&\partialder[L']{q} - \der[]{t}\bracket{\partialder[L']{\dot{q}}} + \partialder[L']{S'} \partialder[L']{\dot{q}} = 0
\end{align}
In the case of $L$ independent of $S$ (i.e. the original Lagrange formalism) one can select $S'(q,t,S)$ in the form
\begin{equation}
S'=S+f(q,t)
\end{equation}
Then eq. \ref{16} reduces to the standard formula. However, even in this case one can put
\begin{equation}
S'=\lambda S
\end{equation}
which leads to the rescaling of the Lagrange function.

The great advantage of the Lagrangian formalism is its invariance under arbitrary invertible and sufficiently smooth transformations on configuration space. The prescription for constructing a new Lagrangian is particularly simple: one has only to express the initial Lagrangian in terms of new coordinates in configuration space. The generalized variational principal shares this useful property. Namely, consider the following general transformation
\begin{align}
\label{21}
t'=t'(t,q,S)\\
\label{22}
q'=q'(t,q,S)\\
\label{23}
S'=S'(t,q,S)
\end{align}
which we assume to be invertible. The reasoning similar to that leading to eq. \ref{16} yields the following formula for the new Lagrangian:
\begin{equation}
\label{24}
L'\bracket{q', \der[q']{t'}, t', S'} \der[t']{t} = \bracket{\partialder[S']{t} + \partialder[S']{q} \der[q]{t} + \partialder[S']{S} L}
\end{equation}
An important consequence of Lagrangian formalism is Noether's theorem relating symmetry transformations and conservation laws. Herglotz formalism leads also to some form of Noether theorem \cite{10,11,12,13} (and the second Noether theorem for gauge symmetries).

The symmetry condition for a given transformation \ref{21}--\ref{23} states that the equations of motion in terms of new variables take the same functional form as those written in terms of initial variables. Therefore, eq. \ref{24} implies the following symmetry condition in terms of Lagrangian function:
\begin{equation}
\label{25}
L\bracket{q', \der[q']{t'}, t', S'} \der[t']{t} =  \bracket{\partialder[S']{t} + \partialder[S']{q} \der[q]{t} + \partialder[S']{S} L}
\end{equation}
Let us note that, contrary to the standard case, we don’t take into account that the final Lagrangian may differ from that on the left-hand side by a counterpart of total derivative. This is because, as it is clearly seen from eqs \ref{16} and \ref{25}, this would only amount to the modification of the function $S'(...)$.

Consider now an infinitesimal form of the transformations \ref{21}--\ref{23}:
\begin{align}
\label{26}
t'=t + \delta t (q,t,S)\\
\label{27}
q'=q +\delta q(q,t,S)\\
\label{28}
S'=S+\delta S(q,t,S)
\end{align}
Inserting \ref{26}--\ref{28} into eq. \ref{25} and expanding to the first order one finds the identity
\begin{gather}
\label{29}
\nonumber
\der[]{t} \bracket{L \delta t + \partialder[L]{\dot{q}} (\delta q - \dot{q} \delta t) -\delta S} - \partialder[L]{S} \bracket{L \delta t + \partialder[L]{\dot{q}}(\delta q -\dot{q} \delta t)-\delta S}\\
+ \bracket{\partialder[L]{q} - \der[]{t} \bracket{\partialder[L]{\dot{q}}} + \partialder[L]{S} \partialder[L]{\dot{q}}} \bracket{\delta q -\dot{q} \delta t}\\
\nonumber
 + (\dot{S} -L) \bracket{\partialder[\delta S]{S} - \partialder[L]{S} \delta t}=0
\end{gather}
On the physical trajectories (cf. eqs.\ref{10}, \ref{11}) eq.\ref{29} can be rewritten as
\begin{equation}
\label{30}
\der[]{t} \bracket{e^{-\int^t \partialder[L]{S} \dt[t']} \bracket{L \delta t + \partialder[L]{\dot{q}}(\delta q - \dot{q} \delta t)-\delta S}} = 0
\end{equation}
The above conservation law is slightly more general than that derived in Ref \cite{10}. The term containing $\delta S$ corresponds to the freedom of adding the total derivative in the standard formalism. There the necessity of including such a term may follow from the cohomological structure of symmetry group \cite{14}. We shall see below that its existence is also necessary in the context of the Fermat principle. 

Contrary to the case of usual Noether theorem, the conservation law \ref{30} is not very useful, at least as it stands. Due to the exponential factor the resulting integral of motion is not a local function of generalized coordinates and velocities; on the contrary, its value depends on whole trajectory in some finite time interval. Let us, however, note that the non-local exponential factor is universal, i.e. it does not depend on the particular form of symmetry transformations. Therefore, given $n$ functionally independent integrals of the form \ref{30} one can construct $n-1$
local independent integrals by taking the quotients. We shall make use of this important property in Sec.5.
 
In analogy with the standard analytic mechanics, one can introduce the counterpart of Hamiltonian formalism \cite{4}. To this end the canonical momentum is introduced
\begin{equation}
p \equiv \partialder[L]{\dot{q}} \equiv p(q, \dot{q}, t,S)
\end{equation}
Assuming non-degeneracy one can compute generalized velocity in terms of momentum, $\dot{q} = \dot{q}(q,p,t,S)$, and define the Hamiltonian
\begin{equation}
\label{32}
H(q,p,t,S) \equiv p \dot{q} (q,p,t,S) - L(q,\dot{q}(q,p,t,S), t,S)
\end{equation}
Eqs. \ref{10}, \ref{11} are equivalent to the generalized canonical equations:
\begin{align}
\label{33}
\dot{q} = \partialder[H]{p}\\
\dot{p} = - \partialder[H]{q} - p \partialder[H]{S}\\
\label{35}
\dot{S} = p \partialder[H]{p} - H
\end{align}
The notion of Poisson bracket is replaced by that of Meyer bracket $(f=f(q,p,t,S)$, $g=g(q,p,t,S))$,
\begin{equation}
\{ f,g\} \equiv \partialder[f]{q} \partialder[g]{p} - \partialder[f]{p} \partialder[g]{q} + p \bracket{\partialder[f]{S} \partialder[g]{p} - \partialder[f]{p} \partialder[g]{S}}
\end{equation}
enjoying the properties:
\begin{itemize}
\item
\begin{equation}
 \{f,g\} = - \{g,f\}
\end{equation}
(antisymmetry)
\item 
\begin{equation}
\{ \alpha f +\beta g, h\} = \alpha \{f,h\} + \beta \{g,h\}
\end{equation}
(linearity)
\item 
\begin{equation}
\{fg,h\} = f \{g,h\} + \{f,h\} g
\end{equation} 
(Leibnitz property)
\item 
\begin{equation}
\begin{aligned}
\left\{ f, \{g,h\} \right\} + \left\{ g,\{h,f\} \right\} + \left\{ h,\{f,g\} \right\}\\
 + \partialder[f]{S} \{g,h\} + \partialder[g]{S} \{h,f\} + \partialder[h]{S} \{f,g\} = 0
\end{aligned}
\end{equation}
(Jacobi identity) 
\end{itemize}
Eqs.\ref{33}-\ref{35} are equivalent to the following ones valid for any (differentiable) function f:
\begin{equation}
\label{41}
\dot{f} = \{f,H\} - H \partialder[f]{S} + \partialder[f]{t}
\end{equation}
Eq. \ref{41}, together with the Jacobi identity, implies
\begin{equation}
\label{42}
\der[]{t} \{f,g\} = \left\{ \der[f]{t}, g \right\} + \left\{ f, \der[g]{t} \right\} + \partialder[H]{S} \{f,g\} 
\end{equation}
It is not difficult to conclude from eq. \ref{42} that
\begin{equation}
\label{43}
\{f,g\}_t = \rho \{f,g\}_0
\end{equation}
with
\begin{equation}
\label{44}
\rho = \exp\bracket{\int_0^t \partialder[H]{S} \dt'} = \exp \bracket{-\int^t_0\partialder[L]{S} \dt'}
\end{equation}
where we have used $\partialder[H]{S} = - \partialder[L]{S}$. The left-hand side of eq. \ref{43} is computed with respect to the canonical variables taken at the time $t$ while on the right side --- at the time $0$.

Herglotz formalism admits natural generalization of the notion of canonical transformation. The transformation
\begin{align}
\label{45}
q' = q'(q,p,S,t)\\
\label{46}
p' = p'(q,p,S,t)\\
\label{47}
S' = S'(q,p,S,t)
\end{align}
is called a canonical one if there exists a function
\begin{equation}
H' = H'(q', p', S', t)
\end{equation}
such that the equations \ref{33}-\ref{35} are equivalent to the same equations in primed variables with $H'$ acting as the
Hamiltonian. It can be shown \cite{4} that \ref{45}-\ref{47} define the canonical transformation provided
\begin{equation}
\label{49}
p' \dt[q'] - \sigma p \dt[q] - (H' - \sigma H) \dt = \dt[S'] - \sigma \dt[S]
\end{equation}
with $\sigma$ being non-vanishing function of canonical variables.

Assuming that \ref{45} can be solved with respect to $p$,
\begin{equation}
p = p(q,q', S,t)
\end{equation}
one can define
\begin{equation}
\tilde{S} (q,q',S,t) = S'(q,p(q,q',S,t),S,t)
\end{equation}
Then \ref{49} may be rewritten as
\begin{equation}
\begin{aligned}
p' \dt[q'] - \sigma p \dt[q] - (H' - \sigma H) \dt = \\ \partialder[\tilde{S}]{q} \dt[q] + \partialder[\tilde{S}]{q'} \dt[q'] + \bracket{\partialder[\tilde{S}]{S} - \sigma} \dt[S] + \partialder[S']{t} \dt.
\end{aligned}
\end{equation}
By comparing the coefficients in front of independent differentials we find
\begin{align}
p = - \frac{1}{\sigma} \partialder[\tilde{S}]{q}\\
\label{54}
p' = \partialder[\tilde{S}]{q'}\\
\sigma = \partialder[\tilde{S}]{S}\\
H'=\sigma H - \partialder[\tilde{S}]{t}
\end{align}
For $H$ independent of $S$ the standard canonical transformations are obtained by putting $\tilde{S} = S + \phi (q,q',t)$.
In many cases, it is more convenient to use the generating function depending on initial coordinate and final momentum. Assume that \ref{46} can be solved with respect to $p$,
\begin{equation}
p = p(q,p',S,t).
\end{equation}
Define the new generating function
\begin{equation}
\label{58}
\psi (q,p',S,t) \equiv p'q'(q,p(q,q',S,t),S,t) - S'(q,p(q,q',S,t),S,t)
\end{equation}
Inserting $S'=p'q'-\psi$ into eq. \ref{49} one finds
\begin{align}
\label{59}
p &= \frac{1}{\sigma} \partialder[\psi]{q}
\\
q'& = \partialder[\psi]{p'}
\\
\label{61}
\sigma &= -\partialder[\psi]{S}
\\
\label{62}
H' &= \sigma H + \partialder[\psi]{t}
\end{align}
Generating function for infinitesimal transformations acquires the form
\begin{equation}
\psi (q,p',S,t) = qp' - S + \delta G(q,p,S,t)
\end{equation}
where $\delta G$, being infinitesimal, can be viewed as a function of initial variables. Putting $q'=q+\delta q$, $p'=p+\delta p$, $S'=S +\delta S$, $\sigma = 1 +\delta \sigma$ we get from \ref{58}-\ref{62}:
\begin{align}
\label{64}
\delta q &= \partialder[\delta G]{p} = \{ q, \delta G\}
\\
\delta p &= -p \partialder[\delta G]{S} - \partialder[ \delta G]{q} = \{ p, \delta G\}
\\
\delta S &= p \partialder[\delta G]{p} - \delta G = \{ S, \delta G \} - \delta G
\\
\label{67}
\delta \sigma &= -\partialder[\delta G]{S}
\end{align}
Using \ref{64}--\ref{67} it is easy to extend the Noether theorem to the case of canonical transformations. The symmetry condition implies that the functional forms of new and old Hamiltonians are the same,
\begin{equation}
\label{68}
H'(q',p',S',t) = H(q',p',S',t)
\end{equation}
From \ref{54} and \ref{64}--\ref{67}, written in infinitesimal form, we find
\begin{equation}
\label{69}
\der[\delta G]{t} + \delta G \partialder[H]{S} = 0
\end{equation}
or
\begin{equation}
\label{70}
\der[]{t}\bracket{\exp\bracket{\int_0^t \partialder[H]{S} \dt'} \delta G}=0
\end{equation}
Noting that $\partialder[H]{S} = -\partialder[L]{S}$ we see that the eq.\ref{70} provides the generalization of eq.\ref{30} to the case of
canonical symmetry transformations. 

Finally, let us derive the Hamilton-Jacobi equation. Putting $H'=0$ in
eq.\ref{62} and using eqs.\ref{59}--\ref{61} we find
\begin{equation}
\partialder[\psi]{t} - \partialder[\psi]{S} H \bracket{q, -\frac{\partialder[\psi]{q}}{\partialder[\psi]{S}}, S,t}=0
\end{equation}
in full agreement with \cite{7}.
\section{Reduced dynamics for homogeneous Lagrangians}

Consider a Lagrangian system with $n+1$ degrees of freedom and generalized coordinates
\begin{equation}
\label{71}
\underline{q} = \bracket{q^0, q^1, ..., q^n} \equiv \bracket{q^\mu},\  \mu=0,1,2,...,n
\end{equation}
Assume the following conditions hold:
\begin{itemize}
\item[(i)] $L$ is non-degenerate,
\begin{equation}
\det \left[ \spartialder[L]{\dot{q}^\mu}{\dot{q}^\nu} \right] \neq 0,\ \dot{q}^\mu \equiv \der[q^\mu]{\sigma};
\end{equation}
\item[(ii)] $L$ does not depend explicitly on the evolution parameter $\sigma$ (time);
\item[(iii)] $L$ is a homogeneous function of generalized velocities of degree 2.
\end{itemize}
The simplest and the most important example of such a Lagrangian is given by
\begin{equation}
\label{73}
L = \frac{1}{2} g_{\mu\nu} (\underline{q}) \dot{q}^\mu \dot{q}^\nu, \ \det[g_{\mu\nu} (\underline{q})] \neq 0
\end{equation}
The trajectories are determined by the Euler-Lagrange equations,
\begin{equation}
\label{74}
\partialder[L]{q^\mu} - \der[]{\sigma} \bracket{\partialder[L]{\dot{q}^\mu}} =0
\end{equation}
It follows from (ii) and (iii) that the energy $E=L$ is conserved:
\begin{equation}
E = E(\underline{q}, \underline{\dot{q}}) \equiv \dot{q}^\mu \partialder[L]{\dot{q}^\mu} - L = 2L - L =L
\end{equation}
is the constant of motion.\\
Let us fix the value $E$ of energy and consider the subset of trajectories corresponding to a given energy value,
\begin{equation}
\label{76}
L(\underline{q},\underline{\dot{q}})=E
\end{equation}
(i) implies that $\partialder[L]{\dot{q}^0} \not\equiv 0$. Therefore, eq. \ref{76} can be solved, at least locally, with respect to $\dot{q}^0$,
\begin{equation}
\label{77}
\dot{q}^0 = \dot{q}^0(q^0, \vec{q}, \dot{\vec{q}}, E),\ \vec{q} \equiv (q^1, ..., q^n)
\end{equation}
implying the identity
\begin{equation}
\label{78}
L(q^0, \vec{q}, \dot{q}^0(q^0, \vec{q}, \dot{\vec{q}};E),\vec{q}) \equiv E
\end{equation}
Differentiating \ref{78} with respect to $q^i$, $\dot{q}^i$ and $q^0$ one finds
\begin{align}
\partialder[\dot{q}^0]{q^i} = - \partialder[L]{q^i} \bracket{\partialder[L]{\dot{q}^0}}^{-1}\\
\partialder[\dot{q}^0]{\dot{q}^i} = - \partialder[L]{\dot{q}^i} \bracket{\partialder[L]{\dot{q}^0}}^{-1}\\
\partialder[\dot{q}^0]{q^0} = - \partialder[L]{q^0} \bracket{\partialder[L]{\dot{q}^0}}^{-1}\\
\end{align}
Let us view the generalized coordinate $q^0$ as the new action variable $S$,
\begin{equation}
q^0 = S
\end{equation}
and define the new action-dependent Lagrangian
\begin{equation}
\label{83}
\lagr(\vec{q}, \dot{\vec{q}}, S;E) \equiv \dot{q}^0 (\vec{q}, \dot{\vec{q}}, q^0; E)
\end{equation}
Eqs. \ref{77}--\ref{83} yield
\begin{equation}
\label{84}
\begin{aligned}
\partialder[\lagr]{q^i} - \der[]{\sigma} \bracket{\partialder[\lagr]{\dot{q}^i}} + \partialder[\lagr]{S} \partialder[\lagr]{\dot{q}^i}\\
= \bracket{\partialder[L]{\dot{q}^0}}^{-1} \bracket{\der[]{\sigma} \bracket{\partialder[L]{\dot{q}^i}} - \partialder[L]{q^i}}\\
 + \bracket{\partialder[L]{\dot{q}^0}}^{-2} \bracket{\partialder[L]{\dot{q}^0}-\der[]{\sigma}\bracket{\partialder[L]{\dot{q}^0}}}\partialder[L]{\dot{q}^i}
\end{aligned}
\end{equation}
It follows then from equations \ref{77}, \ref{83} and \ref{84} that the initial Lagrange equations \ref{74} imply
\begin{align}
\label{85}
&\der[S]{\sigma} = \lagr (\vec{q}, \dot{\vec{q}}, S;E)\\
\label{86}
&\partialder[\lagr]{q^i} - \der[]{\sigma} \bracket{\partialder[\lagr]{\dot{q}^i}} + \partialder[\lagr]{S} \partialder[\lagr]{\dot{q}^i} = 0,\ i=1,2,...,n 
\end{align}
We conclude, therefore, that the initial trajectories, corresponding to a fixed energy $E$, when projected on
coordinates $q^1,...,q^n$, are described by Herglotz variational principle with the action-dependent Lagrangian given by eq.\ref{83}. 

Conversely, assume that the eqs. \ref{85}, \ref{86} admit unique solution obeying the initial condition
\begin{align}
q^i (\sigma_0) &= q^i_0\\
\dot{q}^i(\sigma_0)&=\dot{q}^i_0\\
S(\sigma_0)&=S_0 = q_0^0
\end{align}
Completing the initial conditions by
\begin{equation}
\dot{q}^0(\sigma_0) \equiv \left. \der[q^0]{\sigma} \right|_{\sigma=\sigma_0} = \lagr (\vec{q}_0, \dot{\vec{q}}_0, S_0;E)
\end{equation}
one can find (unique by (i)) solution to the Lagrange equations \ref{74}. Its projection on coordinates $q^1,...,q^n$ yields the
solution we have started with. 

Note that if the $q^0$ variable is cyclic, the resulting reduced dynamics becomes the standard Lagrangian one.

\section{Fermat principle in the presence of gravity}

According to general relativity, the gravitational field is related to the space-time geometry defined by length element
\begin{equation}
\dt[s]^2 = g_{\mu\nu} \dt[x]^\mu \dt[x]^\nu
\end{equation}
In particular, the gravitational field is constant if one can choose a reference frame such that all components $g_{\mu\nu}(x)$ do not depend on $x^0$. If, additionally, $g_{0i}(x)=0$, $i=1,2,3$, the field is called static; in the opposite case, we speak about stationary field.

Light propagation in gravitational field is described by the solutions to Maxwell equations in curved space-time. In the short wavelength approximation one can refer to the geometric optics with its basic notion of light ray. 

In the flat space-time, the light trajectories are determined by Fermat's principle. Its generalization to the case of constant gravitational field is relatively simple. For static fields, it has been
considered in \cite{15,16,17} and extended to stationary ones in \cite{18,19,20}. Particularly elegant derivation of Fermat's
principle for constant gravitational fields (both static and stationary) were given by Landau and Lifshitz \cite{20}.
The starting point is the Fermat principle for flat space-time,
\begin{equation}
\label{92}
\delta \int k_i \dt[x]^i = 0;
\end{equation}
here $k_i$ are covariant components of wave vector, and the integral runs along the light ray trajectory. The
equivalence principle allows us to conclude that \ref{92} is valid also in the presence of gravitational field;
the integrand should be expressed in terms of constant (as measured in the units of $x^0$ along the ray) frequency. This results in the following form of Fermat's principle:
\begin{equation}
\label{93}
\delta \int\bracket{\frac{\dt[l]}{\sqrt{g_{00}}} - \frac{g_{0i}}{g_{00}} \dt[x]^i} =0
\end{equation}
where
\begin{equation}
\dt[l]^2 \equiv \gamma_{ij} \dt[x]^i \dt[x]^j = \bracket{-g_{ij} + \frac{g_{0i} g_{0j}}{g_{00}}} \dt[x]^i \dt[x]^j
\end{equation}
is the space metric.

Alternative method of deriving the equation for light ray trajectory in the presence of gravitational field
is to apply the equivalent principle to the following property of light ray in the flat space-time: the wave
(four)vector is constant along trajectory \cite{20}. As a result, we obtain the geodesic equation
\begin{equation}
\label{95}
\sder[x^\mu]{\sigma} + \Gamma^\mu_{\alpha \beta} \der[x^\alpha]{\sigma} \der[x^\beta]{\sigma} = 0
\end{equation}
with $\sigma$ being an appropriately chosen parameter (the affine parameter). Eq. \ref{95} can be obtained
from variational principle
\begin{equation}
\label{96}
\delta \int \frac{1}{2} g_{\mu \nu} (x) \der[x^\mu]{\sigma} \der[x^\nu]{\sigma} \dt[\sigma] = 0
\end{equation}
supplemented by the condition
\begin{equation}
\label{97}
g_{\mu \nu} (x) \der[x^\mu]{\sigma} \der[x^\nu]{\sigma} = 0
\end{equation}
expressing the fact that the wave fourvector is a null one.

Eqs. \ref{96}, \ref{97} lead to the same trajectories as Fermat's principle. On the other hand,
contrary to the latter, they do not require the assumption concerning the static character of
gravitational field. One can pose the question if they can serve as a starting point for derivation of
Fermat's principle in arbitrary gravitational field.

Let us come back to the static case. Note that solving
\begin{equation}
\label{98}
\dt[s]^2 = 0
\end{equation}
with respect to $\dt[x^0]$ we find
\begin{equation}
\label{99}
\dt[x]^0 = - \frac{g_{0i}}{g_{00}} \dt[x]^i \pm \frac{\dt[l]}{\sqrt{g_{00}}}
\end{equation}
Choosing the upper sign, we see that the Fermat's principal \ref{93} may be rewritten in form
\begin{equation}
\label{100}
\delta \int \dt[x]^0 = 0
\end{equation}
The latter becomes the standard variational principle of Lagrangian mechanics,
\begin{equation}
\label{101}
\delta \int \lagr \dt[\sigma] = 0
\end{equation}
provided we define
\begin{equation}
\label{102}
\lagr \equiv \lagr \bracket{\vec{x}, \der[\vec{x}]{\sigma}} \equiv \der[x^0]{\sigma} 
 = - \frac{g_{0i}}{g_{00}} \der[x^i]{\sigma} + \frac{1}{\sqrt{g_{00}}} \der[l]{\sigma}
\end{equation}
with $\vec{x} \equiv (x^1, x^2, x^3)$. Note that due to the assumption concerning constant gravitational field $\lagr$ does not depend on $x^0$ (see also \cite{101} for Hamiltonian formalism).

The action principle \ref{101} implies the Lagrange equations determining light ray trajectories:
\begin{equation}
\partialder[\lagr]{x^i} - \der[]{\sigma} \bracket{\partialder[\lagr]{\dot{x}^i}} = 0,\ \dot{x}^i \equiv \der[x^i]{\sigma}
\end{equation}
Let us make now the following remarks:
\begin{itemize}
\item[(i)] $\lagr$, as given by equation \ref{102}, is a homogeneous function of first order of the velocities $\dot{x}^i$; therefore,
$\sigma$ can be an arbitrary parametrization of trajectory, not necessarily the initial affine parameter, entering the four-dimensional variational principle \ref{96}. In particular, it is often convenient to choose (at least locally) one of the coordinates $x^i$ as $\sigma$, thereby reducing the problem to two-dimensional dynamics;
\item[(ii)] there are two solutions \ref{99} to the equation \ref{98}. In order to determine the space trajectory we need the initial conditions: initial coordinates and the initial tangent vector. On the other hand, the solution to eq.\ref{95} is determined by the initial point in space-time and initial tangent null fourvector. Knowing the tangent vector to space trajectory, one can determine the zeroth component of tangent null fourvector from the equation $\bracket{\der[s]{\sigma}}^2 = 0$. It has two solutions for $\der[x^0]{\sigma}$ so we have two geodesics with the projections on $x^1, x^2, x^3$ given by the initial conditions for space trajectories and Lagrange equations corresponding to the Lagrangians
\begin{equation}
\label{109}
\lagr_{\pm} = \pm \frac{1}{\sqrt{g_{00}}} \der[l]{\sigma} - \frac{g_{0i}}{g_{00}} \der[x^i]{\sigma}
\end{equation}
It is easy to understand the meaning of the particular choice of sign in \ref{109}. Rewriting eq.\ref{99} in the form
\begin{equation}
\dt[x]^0 + \frac{g_{0i}}{g_{00}} \dt[x^i] = \pm \frac{1}{\sqrt{g_{00}}} \dt[l]
\end{equation}
and using the results concerning the clock synchronization in general relativity \cite{20}, we conclude that the $\lagr_+$ ($\lagr_-$) choice implies the choice of geodesic propagating into future (past), according to the clocks synchronized along the geodesics.
\end{itemize}
Let us note that \ref{100}--\ref{102} imply that $x^0$ plays the role of action $S$ in our Lagrange formalism.

In the case of arbitrary gravitational field eq.\ref{102} takes the form
\begin{equation}
\der[x^0]{\sigma} \equiv \lagr \bracket{\vec{x}, \der[\vec{x}]{\sigma}, x^0} = \frac{1}{\sqrt{g_{00}}} \der[l]{\sigma} - \frac{g_{0i}}{g_{00}} \der[x^i]{\sigma}
\end{equation}
\label{111}
while \ref{95}--\ref{97} are still valid. We can apply the result obtained in the previous sections. Eq.\ref{96} may
be rewritten as
\begin{equation}
\delta \int L \dt[\sigma] = 0
\end{equation}
with
\begin{equation}
L \equiv \frac{1}{2} g_{\mu\nu} (x) \der[x^\mu]{\sigma} \der[x^\nu]{\sigma}
\end{equation}
being homogeneous of degree two in generalized velocities. Therefore, with $\lagr$ defined by eq.\ref{111} and $x^0$ identified with the action variable $S$, the projection of null geodesic on space variables $\vec{x}$ is determined by the Herglotz equations:
\begin{align}
\label{114}
\der[x^0]{\sigma} = \lagr \bracket{\vec{x}, \der[\vec{x}]{\sigma}, x^0}\\
\label{115}
\partialder[\lagr]{\vec{x}} - \der[]{\sigma} \bracket{\partialder[\lagr]{\dot{\vec{x}}}} + \partialder[\lagr]{x^0} \partialder[\lagr]{\dot{\vec{x}}} = 0
\end{align}
Concluding, the Fermat principle for general gravitational field can be formulated as follows: given any
curve $\vec{x} = \vec{x} (\sigma)$, $\sigma_0 \leq \sigma \leq \sigma_1$, $\vec{x}(\sigma_0) = \vec{x}_{(i)}$, $\vec{x}(\sigma_1) = \vec{x}_{(f)}$, one solves eq.\ref{114} subject to the initial condition $x^0(\sigma_0) = x^0_{(i)}$. The physical trajectory $\vec{x}=\vec{x} (\sigma)$ corresponds
to the stationary value of $x^0(\sigma_1)$.

Alternative formulation of Fermat's principle has been proposed in \cite{21,22,23,24}. It can be stated as follows \cite{21,22}.
Assume that the light is emitted at the point p of space-time and received by an observer O at some point q on his worldline $\lambda$. Consider the set N of null curves originating from p and reaching $\lambda$.
The light ray is characterized by the property that its arrival on $\lambda$ is stationary with respect to first order variations of null curve within N. In more detail, let us choose an arbitrary
parameterization $\tau$ of the worldline $\lambda$. To any curve belonging to N there corresponds definite value of the parameter $\tau$. The light ray trajectory corresponds to the curve, for which the value $\tau$ is stationary with respect to the variations of the curve which belong to N. In other words, the null curves belonging to the infinitesimal neighbourhood of p and intersecting $\lambda$, intersect the
latter at the same point q. The reduced curves corresponding to the null curves from this neighbourhood, originate from $\vec{x}_p$ and terminate at $\vec{x}_q$ (cf. Fig 1).

\begin{figure}
\centering
\begin{tikzpicture}[scale=1]
\begin{scope}[xscale = 1,yscale = 1]

\draw[->] (0,0,0) -- (7,0,0) node[anchor=north east]{};
\draw[->] (0,0,0) -- (0,5,0) node[anchor=north west]{$x_0$};
\draw[->] (0,0,0) -- (0,0,5) node[anchor=south]{$\vec{x}$};

\draw (6, 3, 1) .. controls (3.5, 3, 0)  .. (2,2,1);
\draw[line width=0.5mm] (6, 3, 1) .. controls (4, 3, 1)  .. (2,2,1);
\draw (6, 3, 1) .. controls (4, 2.5, 1)  .. (2,2,1);
\filldraw[black] (6,3,1) circle (2pt) node[anchor=south] {q};
\filldraw[black] (2,2,1) circle (2pt) node[anchor=south] {p};
\filldraw[black] (6,0,1) circle (2pt) node[anchor=north] {$\vec{x}_q$};
\filldraw[black] (2,-1,1) circle (2pt) node[anchor=north] {$\vec{x}_p$};

\draw (6, 0, 1) .. controls (3.5, 0, 0)  .. (2,-1,1);
\draw[line width=0.5mm] (6, 0, 1) .. controls (4, 0, 1)  .. (2,-1,1);
\draw (6, 0, 1) .. controls (4, -0.5, 1)  .. (2,-1,1);

\draw[dashed] (6,0,1) -- (6,3,1);
\draw[dashed] (2,-1,1) -- (2,2,1);
\node at (7,4,-1)[above] {$\lambda$};
\draw (7,4,-2) to[out=-120, in=90, looseness=1] (5,1,2);
\end{scope}
\end{tikzpicture}
\caption{The geometry of deflection angle.} \label{fig1}
\end{figure}
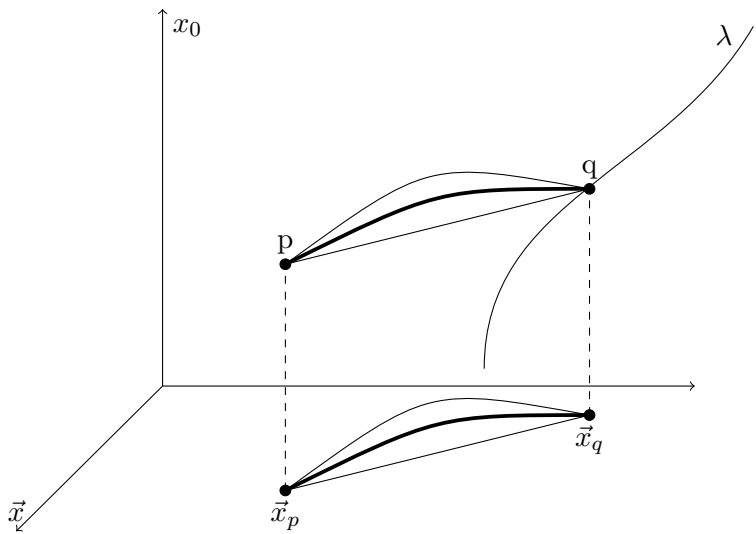

Conversely, consider the reduced curves belonging to the infinitesimal neighbourhood of the reduced
trajectory of light ray. They can be lifted to the null curves originating from p. Since the action $S$ is identified with the coordinate $x^0$, by virtue of Herglotz variational principle, $x^0$ is stationary with respect to the variations of the reduced trajectory. Therefore, their null lifts terminate at q. We conclude that the Fermat principle
based on Herglotz variational principle agrees with the alternative formulation proposed earlier.

There exists still another approach to Fermat's principle based on Pontryagin's minimum principle \cite{25}. The
latter is related to Herglotz formalism \cite{7}, so the results obtained in \cite{25} can be derived within our framework.

As it is had been noticed in \cite{22}, the motion of massive particles can be also described by a counterpart of
Fermat's principle. This may be also easily concluded from the algorithm presented in Section 3. Space-time trajectories of massive particles are described by the solutions to the geodesic equations obeying
\begin{equation}
g_{\mu \nu} \der[x^\mu]{\sigma} \der[x^\nu]{\sigma} > 0
\end{equation}
In this case, the affine parameter is a linear function of $s$; taking $s=\sigma$ we find
\begin{equation}
g_{\mu \nu} (x) \der[x^\mu ]{\sigma} \der[x^\nu]{\sigma} =1
\end{equation}
The corresponding Lagrange function takes the form
\begin{equation}
\label{118}
\lagr_{\pm} \bracket{\vec{x}, \der[\vec{x}]{S}, x^0} \equiv \der[x^0]{s} = - \frac{g_{0i}}{g_{00}} \der[x^i]{s} \pm \frac{\sqrt{\bracket{\der[l]{s}}^2}}{\sqrt{g_{00}}}
\end{equation}
Like in the massless case, the choice of the sign corresponds to propagation forward or backward in time. On the other hand, contrary to the case of null geodesics, there is no gauge symmetry corresponding to
reparameterization invariance.

As we have learned in Section 2, the Herglotz principle admits Hamiltonian formulation. Consider first the case of massive particle propagation, which is simpler because of the lack of reparametrization invariance and consequently, the non-degeneracy of the Lagrangian. Eq.\ref{118} implies (with the choice of $\lagr = \lagr_+$)
\begin{equation}
\label{119}
p_i \equiv \partialder[\lagr]{\bracket{\der[x^i]{s}}} = - \frac{g_{0i}}{g_{00}} + \frac{1}{\sqrt{g_{00}}} \frac{\gamma_{ij} \der[x^j]{s}}{\sqrt{\bracket{\der[l]{s}}^2 +1}}
\end{equation}
One can solve \ref{119} with respect to $\der[x^i]{s}$:
\begin{equation}
\der[x^i]{s} = \gamma^{ij} \Pi_j \frac{\sqrt{g_{00}}}{\sqrt{1-g_{00} \Pi^2}} = - g^{ij} \Pi_j \frac{\sqrt{g_{00}}}{\sqrt{1-g_{00} \Pi^2}}
\end{equation}
where $\gamma^{ij} = - g^{ij}$ \cite{20} and
\begin{align}
\label{121}
\Pi_i &\equiv p_i + \frac{g_{0i}}{g_{00}}
\\
\Pi^2 &\equiv \gamma^{ij} \Pi_i \Pi_j = -g^{ij} \Pi_i \Pi_j 
\end{align}
The relevant Hamiltonian reads:
\begin{equation}
\ham \equiv p_i \der[x^i]{S} - \lagr = - \frac{\sqrt{1-g_{00} \Pi^2}}{\sqrt{g_{00}}}
\end{equation}
According to \ref{33}--\ref{35} the equations describing the particle trajectory read:
\begin{align}
\label{124}
\der[x^i]{s} &= \partialder[\ham]{p_i}\\
\label{125}
\der[p_i]{s} &=- \partialder[\ham]{x^i} - p_i \partialder[\ham]{x^0}\\
\label{126}
\der[x^0]{s} &= p_i \partialder[\ham]{p_i} - \ham
\end{align}
The case of massless trajectories (light rays) is more complicated. The Lagrangian \ref{109} (again we choose $\lagr \equiv \lagr_+$) and
eqs. \ref{114}, \ref{115} define the reparameterization invariant dynamics; $\lagr$ is degenerate. The generalized momenta take the form
\begin{equation}
\label{127}
p_i = - \frac{g_{0i}}{g_{00}} + \frac{1}{\sqrt{g_{00}}} \frac{\gamma_{ij} \der[x^j]{\sigma}}{\der[l]{\sigma}}
\end{equation}
Eqs. \ref{127} imply the following constraint:
\begin{equation}
\label{128}
g_{00} \gamma^{ij} \Pi_i \Pi_j -1 =0
\end{equation}
with $\Pi_i$ defined by eq. \ref{121}. Due to the constraint \ref{128} eq.\ref{127} cannot be solved with respect to $\der[x^i]{s}$. Moreover, the reparameterization invariance implies the vanishing of canonical
Hamiltonian
\begin{equation}
\ham \equiv \dot{x}^i \partialder[\lagr]{\dot{x}^i} - \lagr \equiv 0
\end{equation}
Therefore, the construction of Hamiltonian formalism requires the extension of Dirac formalism \cite{102}. By analyzing the Herglotz version of Hamiltonian formalism for non-degenerate case one can conclude that its generalization to the degenerate case may be quite straightforward. However, we are here interested in
the particular gauge symmetry - reparametrization invariance. Therefore, we assume that the correct Hamiltonian is simply
proportional to the constraint \ref{128},
\begin{equation}
\ham = \mu \bracket{g_{00} \gamma^{ij} \Pi_i \Pi_j -1},
\end{equation}
with $\mu$ being the relevant Lagrange multiplier. Eqs. \ref{124}--\ref{126} take then the form:
\begin{align}
\label{131}
\dot{x}^i =&\ 2 \mu g_{00} \gamma^{ij} \Pi_j
\\
\label{132}
\dot{p}_i = &-\mu\ \partial_i (g_{00} \gamma^{kl})\ \Pi_k\ \Pi_l - 2 \mu\ \gamma_{00}\ \gamma^{kl} \partial_i g_k\ \Pi_l
\\
\nonumber
&-\mu (\Pi_i - g_i )\ \partial_0 (g_{00} \gamma^{kl})\ \Pi_k\  \Pi_l - 2 \mu (\Pi_i - g_i ) \gamma^{kl}\ \partial_0 g_k\ \Pi_l
\\
\label{133}
\dot{x}^0 = &\mu \bracket{g_{00}\ \gamma^{kl} (p_k - g_k)\ \Pi_l +1}
\end{align}
where
\begin{equation}
g_i \equiv \frac{g_{0i}}{g_{00}}
\end{equation}
In particular, equation \ref{131} yields
\begin{equation}
\label{135}
\Pi_i = \frac{\gamma_{ij} \dot{x}^j}{2 \mu g_{00}}
\end{equation}
Eq.\ref{135} and the constraint \ref{128} imply
\begin{equation}
\label{136}
\mu = \frac{1}{2} \sqrt{\frac{\gamma_{ij} \dot{x}^i \dot{x}^j}{g_{00}}}
\end{equation}
(second solution, differing by sign, leads to the same conclusions).
Differentiating \ref{131} over $\sigma$ and using \ref{132}, \ref{133} and \ref{136} we conclude that the resulting equations are equivalent to the Herglotz ones,
\begin{align}
\dot{x}^0 = \lagr\\
\partialder[\lagr]{x^i} - \der[]{\sigma} \bracket{\partialder[\lagr]{\dot{x}^i}} + \partialder[\lagr]{x^0} \partialder[\lagr]{\dot{x}^i} = 0
\end{align}
The actual computations are slightly involved and are omitted here. We see that the Dirac formalism can be
directly generalized, at least in the case of reparameterization invariant theory.

An alternative approach to the Hamiltonian formalism relies on the gauge fixing procedure. Particularly convenient gauge condition is obtained by identifying (at least locally) the evolution parameter $\sigma$ with one of the space coordinates, for example $\sigma=x^3$. Then the relevant Lagrangian takes the form
\begin{equation}
\lagr = - \frac{g_{0a} \dot{x}^a}{g_{00}} - \frac{g_{03}}{g_{00}} + \frac{1}{\sqrt{g_{00}}} \sqrt{\gamma_{ab} \dot{x}^a \dot{x}^b + \gamma_{a3} \dot{x}^a + \gamma_{33}} 
\end{equation}
where that summation runs over $a,b=1,2$.
The Lagrange equations take the form:
\begin{align}
&\partialder[\lagr]{x^a} - \der[]{\sigma} \bracket{\partialder[\lagr]{\dot{x}^a}} + \partialder[\lagr]{x^0} \partialder[\lagr]{\dot{x}^a} = 0,\ a=1,2
\\
&\dot{x}^0 = \lagr
\\
&\sigma = x^3
\end{align}
Let us point out that in the massless case, due to the reparametrization invariance, Fermat's principle reduces the
number of dependent variables to three, $x^1$, $x^2$ and $x^0$.

\section{Fermat principle for Schwarzshield-deSitter metric}

As an example, we shall consider the Schwarzshield-deSitter metric. It is the spherically symmetric metric describing a black
hole in the space-time with non-vanishing cosmological constant. The relevant line element reads \cite{26}
\begin{align}
\label{143}
\dt[s]^2 &= B(R) \dt[T]^2 - \frac{\dt[R^2]}{B(R)} - R^2 \bracket{\dt[\theta]^2 + \sin^2 \theta \dt[\phi]^2}\\
\label{144}
B(R) &= 1 - \frac{2m}{R} - \frac{\Lambda}{3} R^2
\end{align}
and $\Lambda>0$ (for $\Lambda<0$ we are dealing with Schwarzshield-anti-deSitter metric). Assuming that
\begin{equation}
0< \Lambda < \frac{1}{9} m^2
\end{equation}
we conclude \cite{27}, \cite{28} that the equation
\begin{equation}
B(R) = 0
\end{equation}
has three real roots
\begin{equation}
R_1 < 0 < R_2 < R_3
\end{equation}
and $B(R)>0$ for $R_2 < R < R_3$. Therefore, the region between the horizons $R=R_2$ and $R=R_3$ corresponds to the static metric. For the remaining $\Lambda>0$ the static region does not exist \cite{27,28}.

It is well known that the geodesic equations for spherically symmetric metric are integrable in the Arnold-Liouville sense.

Performing the coordinate transformation
\begin{align}
\label{148}
R = a(t) r (1+\mu)^2\\
T= t + \zeta(R)
\end{align}
with
\begin{align}
\label{150}
a(t) &\equiv e^{\sqrt{\frac{\Lambda}{3}} t } \equiv e^{Ht}\\
\label{151}
\mu &\equiv \frac{m}{2a(t)r}\\
\label{152}
\der[\zeta]{R} &= \frac{HR}{\sqrt{1-\frac{2m}{R}} \bracket{1-\frac{2m}{R} - \frac{\Lambda}{3} R^2}}
\end{align}
one obtains McVittie metric \cite{29,28}
\begin{equation}
\label{153}
\dt[s]^2 = \bracket{\frac{1-\mu }{1+\mu }}^2 \dt - (1+\mu )^4 a^2 \bracket{\dt[r]^2 + r^2 (\dt[\theta]^2 + \sin^2 \theta \dt[\phi]^2)} 
\end{equation}
For $m=0$ we find that \ref{153} reduces to the Friedman--Lemaitre-- Robertson--Walker (FLRW) metric. On the other hand, for $\Lambda=0$ we obtain
\begin{equation}
\label{154}
\dt[s]^2 = \bracket{\frac{1-\frac{m}{2r}}{1+\frac{m}{2r}}}^2 \dt^2 - \bracket{1 + \frac{m}{2r}}^4 \bracket{\dt[r]^2 + r^2 (\dt[\theta]^2 + \sin^2 \theta \dt[\phi]^2)} 
\end{equation}
which, when expressed in terms of $\rho \equiv r \bracket{1+\frac{m}{2r}}^2$, reduces to the Schwarzshield metric.

In order to describe the propagation of light in terms of coordinates entering \ref{153} we resort to the
Fermat principle. The metrics depends now explicitly on time, so one has to use Fermat's principle in Herglotz form. Due to the rotational invariance one can assume $\theta = \frac{\pi}{\theta}$ from the very beginning. The reparametrization invariance allows, in turn, to select $\phi$ as the evolution parameter. Solving $\dt[s]^2=0$ we find from eq.\ref{153}:
\begin{equation}
\lagr (r,\dot{r}, t) \equiv \der[t]{\phi} = \frac{(1+\mu)^3}{1-\mu} a(t) \sqrt{\dot{r}^2 + r^2},\ \dot{r} \equiv \der[r]{\phi}
\end{equation}
where $t (\equiv S)$ plays the role of action variable. 

The Herglotz equations
\begin{align}
\label{156}
\der[t]{\phi} = \lagr (r,\dot{r}, t)\\
\partialder[\lagr]{r} - \der[]{\phi} \bracket{\partialder[\lagr]{\dot{r}}} + \partialder[\lagr]{t} \partialder[\lagr]{\dot{r}} = 0
\end{align}
yield
\begin{align}
\label{158}
&\dot{t} = h \sqrt{\dot{r}^2 +r^2}\\
\label{159}
&r \ddot{r} - 2\dot{r}^2 - r^2 - \partialder[(\ln h)]{r}\ r\ (\dot{r}^2 + r^2) = 0
\end{align}
The generalized Noether's theorem described in Section 2 may be stated as follows. Assume that
\begin{align}
r &\to r + \delta r \\
t &\to t + \delta t \\
\phi &\to \phi + \delta \phi
\end{align}
is the symmetry transformation (in the sense defined there); then
\begin{equation}
\label{163}
C = \exp\bracket{- \int_0^\phi \partialder[\lagr]{t} \dt[\phi]} \bracket{\delta \phi \lagr + ( \delta r - \dot{r} \delta \phi) \partialder[\lagr]{\dot{r}} - \delta t} 
\end{equation}
is an integral of motion. As we have already mentioned, the integral \ref{163} is not very useful. Due to
the exponential factor it is non-local, i.e. it depends on the whole trajectory in the interval $\left< 0, \phi \right>$. However, due to
the universality of non-local factor one can obtain an integral of motion from two functionally
independent symmetry transformations.
First, note that $\lagr$ does not depend on $\phi$. So, as in the standard Lagrangian formalism, one can consider the
transformations
\begin{align}
\label{164}
\delta \phi &= \psi \ (\equiv \text{constant})\\
\delta r &= 0\\
\label{166}
\delta t &= 0
\end{align}
Eq.\ref{25} is obviously obeyed and the counterpart of energy,
\begin{equation}
\label{167}
C_1 \equiv \exp\bracket{- \int_0^\phi \partialder[\lagr]{t} \dt[\phi]} \bracket{\lagr - \dot{r} \partialder[\lagr]{\dot{r}}}
\end{equation}
which reads
\begin{equation}
\label{168}
C_1 = \exp\bracket{- \int_0^\phi \partialder[h]{t} \sqrt{\dot{r}^2 + r^2} \dt[\phi]} \frac{hr^2}{\sqrt{\dot{r}^2 +r^2}}
\end{equation}
is conserved.

Consider, in turn, the following transformations
\begin{align}
\delta \phi &= 0\\
\delta t &= \tau\  (\equiv \text{constant})\\
\delta r &= - Hr\tau \equiv - \sqrt{\frac{\Lambda}{3} } r \tau
\end{align}
Again, eq.\ref{25} is fulfilled and eq.\ref{163} yields
\begin{equation}
\label{172}
C_2 = \exp\bracket{- \int_0^\phi \partialder[\lagr]{t} \dt[\phi]} \bracket{\lagr - Hr \partialder[\lagr]{\dot{r}} -1}
\end{equation}
Or explicitly
\begin{equation}
C_2 = -\exp\bracket{- \int_0^\phi \partialder[h]{t} \sqrt{\dot{r}^2 + r^2} \dt[\phi]} \frac{Hhr\dot{r}}{\sqrt{\dot{r}^2 +r^2}}+1
\end{equation}
The quotient of $C_1$ and $C_2$ is the local integral of motion
\begin{equation}
C \equiv - \frac{C_2}{C_1} = H \frac{\dot{r}}{r} + \frac{\sqrt{\dot{r}^2+r^2}}{hr^2}
\end{equation}
Using eqs. \ref{158}, \ref{159} it is straightforward to check that $C$ is indeed an integral of motion, $\dot{C}=0$.

Equations \ref{150}, \ref{151} and \ref{156} imply
\begin{equation}
\frac{\dot{\mu}}{\mu} = - \frac{\dot{r}}{r} - Hh\sqrt{\dot{r}^2 +r^2}
\end{equation}
Using the last equation one can show that
\begin{equation}
\label{176}
C^2 + H^2 = \frac{(1-\mu)^2}{(1+\mu)^6} \bracket{\dot{\mu}^2 + \mu^2}
\end{equation}
Concluding, the initial system \ref{158}, \ref{159} reduces to the set of two equations of the first order,
\begin{align}
\label{177}
&\dot{t} = h \sqrt{\dot{r}^2 + r^2}\\
\label{178}
&\frac{H \dot{r}}{r} + \frac{\sqrt{\dot{r}^2 + r^2}}{hr^2} = C
\end{align}
Eqs.\ref{177}, \ref{178} are integrable by quadratures. First note, that the resulting relation \ref{176} is the
first order equation for $\mu$, which is integrable by separating variables yielding the function $\mu = \mu(\phi)$. By defining
\begin{equation}
\rho \equiv \ln r
\end{equation}
we can rewrite equation \ref{178} in the form
\begin{equation}
\label{180}
H \dot{\rho} + F(\mu) \sqrt{\dot{\rho}^2 +1} = C
\end{equation}
with
\begin{equation}
\label{181}
F(\mu) \equiv \frac{2\mu (1-\mu )}{M (1 +\mu^3)}
\end{equation}
being a known function of $\phi$, $\tilde{F}(\phi) \equiv F(\mu(\phi))$. Solving \ref{180} with respect to $\dot{\rho}$ and integrating we find $r = r(\phi)$. Finally, eq.\ref{177} can be written as
\begin{equation}
\label{182}
\der[]{\phi} \bracket{e^{-Ht}} = \frac{-H(1+\mu)^3}{1-\mu} \sqrt{\dot{r}^2 + r^2}
\end{equation}
The function $t=t(\phi)$ is then found by integration of both sides with respect to $\phi$.

Remark that, although we had to use the Herglotz formalism, this is not the most general case. By passing to the static coordinates (cf. eqs. \ref{143}, \ref{144} and \ref{148}--\ref{152}) everything reduces to the
standard Lagrangian formalism. In particular, the integral of motion results from evolution parameter independence in static coordinates.
\section{Conclusions}
We have considered autonomous non-degenerate Lagrangians which are homogeneous of degree 2 in
generalized velocities. From the energy integral one can compute one generalized velocity. The original
dynamics when reduced to the remaining generalized coordinates can be described by the Herglotz variational principle with the distinguished coordinate and velocity playing the role of action variable and action-dependent Lagrangian, respectively.

Some remarks are in order here.
\begin{itemize}
\item[(i)] we do not have to assume that the homogeneity degree equals 2. The non-degeneracy implies it cannot be 1. Assuming $\lagr$ is homogeneous of degree $k\neq1$ we find $E=\dot{q}^i \partialder[\lagr]{\dot{q}^i} - \lagr = (k-1) \lagr$ and one can repeat that reasoning
presented in section 3.;
\item[(ii)]
the initial dynamics involves $n+1$ equations of second degree. After performing that reduction
procedure one obtains, in general, $n$ equations \ref{86} of second degree and one, eq.\ref{85}, of first degree. However, for $E=0$ the resulting Lagrangian $\lagr$, being homogeneous of first degree, is degenerate.
Therefore, we can fix the gauge, say $\sigma=q^n$ obtaining $n-1$ equations of second degree and one of
first degree. 
\end{itemize}
The above simple result allows us to derive the counterpart of Fermat's principle for arbitrary gravitational field.
Starting from the quadratic Lagrangian describing the geodesics in affine parameterization we select the sub-manifold of solutions corresponding to vanishing 'energy' $E=L$. We compute the effective Lagrangian $\lagr$ by solving $\lagr=0$ with respect to $\dot{x}^0$.
Then $\lagr$ becomes an action-dependent Lagrangian with $x^0$ playing the role of action variable.
Moreover, $\lagr$ defines the reparameterization invariant dynamics, so the evolution parameter may be
identified, at least locally, with some coordinate reducing thereby the number of variables involved. 

In this way, the Fermat principle takes the form of Herglotz variational principle. This approach agrees with alternative ones, proposed in \cite{21,22,23,24,25}. Moreover, by considering the reduction procedure for non-vanishing $E$ one can formulate a counterpart of Fermat's principle for propagation of massive particles.
Note also that the reasoning described in Section 3 can be applied to any coordinate. In the context of light propagation, this means that any coordinate can serve as action variable. Taking, for example, $S=x^3$ one
arrives at the form of Fermat's principle describing the evolution of $x^0$, $x^1$ and $x^2$ variables. Finally, let us note,
that the Herglotz variational principle has been considered in the context of general relativity (in application to homogeneous cosmologies) in \cite{36,37}.
\section*{Acknowledgments}
The inspiring discussion with Krzysztof Andrzejewski, Cezary Gonera and Paweł Maślanka are gratefully acknowledged.

\bibliographystyle{ieeetr}
\bibliography{bibliography}

\end{document}